\documentclass[11pt]{amsart}
\usepackage[margin=1.1in]{geometry}
\usepackage{amsmath,amssymb}
\usepackage{enumitem}
\usepackage{xcolor}
\usepackage{tikz}
\usepackage[colorlinks=true,linkcolor=blue!50!black,citecolor=green!40!black,urlcolor=blue!50!black]{hyperref}

\newtheorem{theorem}{Theorem}[section]

\newtheorem{lemma}[theorem]{Lemma}

\newtheorem{claim}{Claim}
\theoremstyle{definition}
\newtheorem{definition}[theorem]{Definition}

\theoremstyle{remark}
\newtheorem{remark}[theorem]{Remark}

\newcommand{\cmu}{\chi_\mu}
\newcommand{\cH}{\mathcal{H}}
\newcommand{\R}{\text{red}}
\newcommand{\B}{\text{blue}}

\title[Mutual-visibility colouring in fixed diameter is hard]{Mutual-visibility $k$-colouring is NP-complete\\ for every fixed diameter}
\author{Siddhesh Chaubal}
\address{Afinit, India}
\email{spchaubal@gmail.com}

\subjclass[2020]{05C15, 05C12, 68Q17}
\keywords{mutual-visibility set, mutual-visibility colouring, diameter two, NP-completeness, hypergraph 2-colouring}

\begin{document}

\begin{abstract}
A \emph{mutual-visibility set} in a graph $G$ is a set $S$ of vertices such that any two vertices of $S$ are connected by a shortest path with no internal vertex in $S$, and the \emph{mutual-visibility chromatic number} $\cmu(G)$ is the minimum number of mutual-visibility sets needed to partition $V(G)$. Babu, Di Stefano and Lakshmanan showed that it is NP-complete to decide whether $\cmu(G)\le 2$ for graphs of diameter four. We show that the problem remains NP-complete for graphs of diameter $d$, for every fixed $d\ge2$; diameter two is the smallest diameter for which the problem is non-trivial. More generally, for all fixed $k\ge2$ and $d\ge2$, it is NP-complete to decide whether $\cmu(G)\le k$ for graphs of diameter $d$. The proof rests on a reformulation of $\cmu(G)\le 2$ for diameter-two graphs as the $2$-colourability of an associated hypergraph, and on a reduction from $2$-colourability of $3$-uniform hypergraphs without short cycles.
\end{abstract}

\maketitle

\section{Introduction}

Mutual visibility in graphs was introduced by Di Stefano~\cite{DiStefano2022}: a set $S\subseteq V(G)$ is called a \emph{mutual-visibility set} (MV set) of $G$ if any two vertices $u,v\in S$ are \emph{$S$-visible}, that is, joined by a shortest $u,v$-path all of whose internal vertices lie outside $S$. Klav\v{z}ar, Kuziak, Valenzuela Tripodoro and Yero~\cite{KKVY2025} introduced the corresponding colouring problem. A partition of $V(G)$ into MV sets is called a \emph{mutual-visibility colouring} of $G$, and the \emph{mutual-visibility chromatic number} $\cmu(G)$ is the minimum number of parts in such a partition. They asked about its computational complexity~\cite[Problem~8.1]{KKVY2025}, and Babu, Di Stefano and Lakshmanan~\cite{BDL2025} proved that it is NP-complete to decide whether $\cmu(G)\le 2$, already for graphs of diameter four. Related variants (independent, total, dual and outer mutual-visibility colourings) are also known to be NP-complete~\cite{BPSY2026,BJKLY2026}.

Graphs of diameter two play a special role for mutual visibility~\cite{CDKY2024}. In such a graph every independent set is an MV set, so $\cmu(G)\le\chi(G)$~\cite{KKVY2025}, and the existence of graphs attaining equality other than complete bipartite graphs was recently settled by the author~\cite{Chaubal}. For graphs of diameter at most one (complete graphs) the problem is trivial. Our main result shows that diameter two is already as hard as the general case.

\begin{theorem}\label{thm:main}
The following problem is NP-complete: given a graph $G$ of diameter two, decide whether $\cmu(G)\le 2$.
\end{theorem}

Since a non-complete graph has $\cmu(G)\ge2$, Theorem~\ref{thm:main} equivalently says that it is NP-complete to decide whether $\cmu(G)=2$ for graphs of diameter two, and computing $\cmu$ is NP-hard on this class. Attaching a suitable `ladder' to the construction extends this to every larger diameter, and in particular recovers the diameter-four case of~\cite{BDL2025}.

\begin{theorem}\label{thm:diamd}
For every fixed integer $d\ge2$, the following problem is NP-complete: given a graph $G$ of diameter $d$, decide whether $\cmu(G)\le 2$.
\end{theorem}

The same reductions, with minor changes, show that deciding $\cmu(G)\le k$ is NP-complete for all fixed $k\ge2$ and $d\ge2$ (Theorem~\ref{thm:kcol} in Section~\ref{sec:kcol}).

The paper is structured as follows. In Section~\ref{sec:reform} we show that, for a graph of diameter two, $\cmu(G)\le 2$ is equivalent to the $2$-colourability of a hypergraph whose edges are the sets $\{u,w\}\cup(N(u)\cap N(w))$ for non-adjacent pairs $u,w$. In Section~\ref{sec:girth} we show that $2$-colourability of $3$-uniform hypergraphs is NP-complete, and, by a standard gadget argument, that this remains true for hypergraphs without Berge cycles of length two or three. In Section~\ref{sec:reduction} we construct, from such a hypergraph $F$, a graph $G(F)$ of diameter two whose associated hypergraph contains every edge of $F$ exactly, while all its other edges can be made harmless; this proves Theorem~\ref{thm:main}. In Section~\ref{sec:diamd} we attach to $G(F)$ a `ladder' of $d-2$ levels, which raises the diameter to exactly $d$ without changing whether $\cmu\le2$; this proves Theorem~\ref{thm:diamd}. In Section~\ref{sec:kcol} we extend both theorems from two colours to any fixed number $k\ge2$ of colours (Theorem~\ref{thm:kcol}).

\section{Preliminaries and a hypergraph reformulation}\label{sec:reform}

All graphs are finite and simple. For a vertex $v$ of a graph $G$, $N(v)$ is its set of neighbours, and for distinct vertices $u,w$ we write $C(u,w)=N(u)\cap N(w)$ for their set of common neighbours; we write $u\sim w$ if $u$ and $w$ are adjacent. A hypergraph is \emph{$3$-uniform} if all its edges have exactly three vertices, and it is \emph{$2$-colourable} if its vertices can be coloured red and blue so that no edge is monochromatic.

Suppose that $G$ has diameter two. Two adjacent vertices are joined by an edge, which is a shortest path with no internal vertex. Two non-adjacent vertices $u,w$ are at distance two, so the shortest $u,w$-paths are the paths $u\,x\,w$ with $x\in C(u,w)$, and $C(u,w)\ne\emptyset$. Consequently, $S\subseteq V(G)$ is an MV set exactly when any two non-adjacent vertices of $S$ have a common neighbour not in $S$.

\begin{definition}
Let $G$ be a graph of diameter two. The \emph{visibility hypergraph} $\cH(G)$ has vertex set $V(G)$ and edge set
\[
\bigl\{\{u,w\}\cup C(u,w)\ :\ u,w\in V(G),\ u\ne w,\ u\not\sim w\bigr\}.
\]
\end{definition}

\begin{lemma}\label{lem:reform}
Let $G$ be a graph of diameter two. Then $\cmu(G)\le2$ if and only if $\cH(G)$ is $2$-colourable.
\end{lemma}

\begin{proof}
A partition of $V(G)$ into at most two MV sets is the same as a colouring $\varphi\colon V(G)\to\{\R,\B\}$ whose two colour classes are MV sets (one of them may be empty). By the observation above, both colour classes are MV sets if and only if any two non-adjacent vertices $u,w$ with the same colour have a common neighbour with the other colour, that is, if and only if no set $\{u,w\}\cup C(u,w)$ with $u\not\sim w$ is monochromatic.
\end{proof}

\section{Hypergraph 2-colouring without short cycles}\label{sec:girth}

A \emph{Berge cycle of length $\ell\ge2$} in a hypergraph consists of distinct vertices $v_1,\dots,v_\ell$ and distinct edges $E_1,\dots,E_\ell$ such that $v_i\in E_i\cap E_{i+1}$ for all $i$ (indices modulo $\ell$). We say that a hypergraph has \emph{girth at least four} if it has no Berge cycle of length two or three; equivalently, any two edges share at most one vertex, and there are no three edges $E_1,E_2,E_3$ and distinct vertices $v_1\in E_1\cap E_2$, $v_2\in E_2\cap E_3$, $v_3\in E_3\cap E_1$.

The NP-completeness of $2$-colourability of hypergraphs goes back to Lov\'asz~\cite{Lovasz1973}; for edges of size at most three, see also~\cite[SP4]{GJ1979}. For completeness we include a short proof for $3$-uniform hypergraphs.

\begin{lemma}\label{lem:3unif}
It is NP-complete to decide whether a $3$-uniform hypergraph is $2$-colourable.
\end{lemma}

\begin{proof}
Membership in NP is clear. We reduce from graph $3$-colourability, which is NP-complete~\cite[GT4]{GJ1979}. Given a graph $G$, let $H$ be the $3$-uniform hypergraph with vertex set $\{x_{i,v}: i\in\{1,2,3\},\ v\in V(G)\}\cup\{z\}$ and the edges
\[
f_v=\{x_{1,v},x_{2,v},x_{3,v}\}\quad(v\in V(G)),\qquad g_{uv,i}=\{z,x_{i,u},x_{i,v}\}\quad(uv\in E(G),\ i\in\{1,2,3\}).
\]
It can be constructed in polynomial time. We show that $G$ is $3$-colourable if and only if $H$ is $2$-colourable.

Let $c$ be a proper $3$-colouring of $G$. Colour $z$ red, and colour $x_{i,v}$ red if $c(v)=i$ and blue otherwise. Each edge $f_v$ contains exactly one red vertex, namely $x_{c(v),v}$. Each edge $g_{uv,i}$ contains the red vertex $z$ and is not entirely red, since $c(u)=c(v)=i$ is impossible for an edge $uv$. So $H$ is $2$-colourable.

Conversely, take a $2$-colouring of $H$; exchanging the colours if necessary, we may assume that $z$ is red. Since $f_v$ is not monochromatic, some $x_{i,v}$ is red; choose such an $i$ and put $c(v)=i$. If $uv\in E(G)$ and $c(u)=c(v)=i$, then the edge $g_{uv,i}$ is entirely red, a contradiction. So $c$ is a proper $3$-colouring of $G$.
\end{proof}

We need the following variant, for hypergraphs without short cycles.

\begin{lemma}\label{lem:girth}
It is NP-complete to decide whether a $3$-uniform hypergraph of girth at least four is $2$-colourable.
\end{lemma}

\begin{proof}
Membership in NP is clear. By a theorem of Erd\H{o}s and Hajnal~\cite[Corollary~13.4]{EH1966}, applied with $k=s=r=3$ (see also~\cite{Lovasz1968,NR1979} for constructive proofs), there is a $3$-uniform hypergraph that is not $2$-colourable and in which any $t\le3$ edges cover at least $2t+1$ vertices. In particular, it has girth at least four, since two edges sharing two vertices cover at most four vertices, and the three edges of a Berge cycle of length three cover at most six. Fix such a hypergraph, and let $H_0$ be an edge-minimal non-$2$-colourable sub-hypergraph of it, that is, $H_0$ is not $2$-colourable but $H_0-e$ is $2$-colourable for every edge $e$ of $H_0$. Such an $H_0$ exists because the hypergraph is finite, and it still has girth at least four. Fix an edge $e_0=\{x,y,z\}$ of $H_0$ and let $D=H_0-e_0$. Then:
\begin{enumerate}[label=(\alph*)]
\item $D$ is $2$-colourable, by the minimality of $H_0$;
\item in every $2$-colouring of $D$ the vertices $x,y,z$ have the same colour, since otherwise the colouring would be a $2$-colouring of $H_0$; together with~(a) and exchanging the colours, $D$ has $2$-colourings in which $x,y,z$ are all red and $2$-colourings in which they are all blue;
\item no edge of $D$ contains two of $x,y,z$, since such an edge would share two vertices with $e_0$ in $H_0$.
\end{enumerate}
The hypergraph $D$ is a fixed object, independent of the input.

We reduce from the problem of Lemma~\ref{lem:3unif}. Let $F$ be a $3$-uniform hypergraph; we may assume that every vertex lies in some edge. For a vertex $v$ of degree $\deg(v)$ in $F$, replace $v$ by $\deg(v)$ new vertices $v^1,\dots,v^{\deg(v)}$, using a different one in each of the $\deg(v)$ edges of $F$ that contain $v$; we call the resulting edges \emph{clause edges}. For every $v$ and every $1\le i<\deg(v)$ add a copy $D_{v,i}$ of $D$ in which $x$ is identified with $v^i$, $y$ with $v^{i+1}$, and all other vertices are new. Let $F'$ be the resulting $3$-uniform hypergraph. Its size is linear in the size of $F$.

\emph{$F$ is $2$-colourable if and only if $F'$ is.} Given a $2$-colouring $\psi$ of $F$, colour every $v^i$ with $\psi(v)$ and extend the colouring to each $D_{v,i}$ by~(b); this is a $2$-colouring of $F'$. Conversely, in a $2$-colouring of $F'$, property~(b) applied to $D_{v,1},\dots,D_{v,\deg(v)-1}$ shows that $v^1,\dots,v^{\deg(v)}$ all have the same colour; giving $v$ this colour yields a $2$-colouring of $F$, because every edge of $F$ corresponds to a clause edge of $F'$.

\emph{$F'$ has girth at least four.} Call the clause edges and the copies $D_{v,i}$ the \emph{pieces} of $F'$; every edge of $F'$ lies in exactly one piece, and we identify a piece with its vertex set. Two clause edges are disjoint; a clause edge and $D_{v,i}$ share at most one vertex (a clause edge contains only one of $v^1,\dots,v^{\deg(v)}$); $D_{v,i}$ and $D_{v,i+1}$ share exactly $v^{i+1}$; all other pairs of pieces are disjoint. In particular, two pieces share at most one vertex.

Two edges in the same piece share at most one vertex, since $D$ has girth at least four, and two edges in different pieces share at most one vertex, since the pieces do. So there is no Berge cycle of length two.

Consider a Berge cycle $v_1,E_1,\dots,v_\ell,E_\ell$. If all $E_i$ lie in one piece, then $\ell\ge4$ because $D$ has girth at least four (and a clause piece has only one edge). Otherwise split the cyclic sequence $E_1,\dots,E_\ell$ into maximal runs of consecutive edges lying in the same piece. Where the cycle passes from a run in a piece $P$ to the next run, in a piece $Q\ne P$, the vertex $v_i$ between them lies in $P\cap Q$. If there were exactly two runs, in pieces $P$ and $Q$, the two transition vertices would both lie in $P\cap Q$, which has at most one element, contradicting their distinctness. Hence there are at least three runs. Consecutive runs lie in pieces that intersect, and two clause edges never intersect, so some run lies in a copy $D_{v,i}$. This run is entered and left through two distinct transition vertices, which are vertices of $D_{v,i}$ shared with other pieces; the only such vertices are $v^i$ and $v^{i+1}$. By~(c) no single edge of $D_{v,i}$ contains both, so this run has at least two edges. There are at least two other runs, with at least one edge each, so $\ell\ge2+1+1=4$.
\end{proof}

\section{The reduction}\label{sec:reduction}

Let $F$ be a $3$-uniform hypergraph of girth at least four with vertex set $X$. For every edge of $F$ choose one of its vertices as its \emph{centre}, and write the edge as $(a,b\mid c)$ with centre $c$. We construct a graph $G=G(F)$ as follows.
\begin{itemize}
\item The \emph{core vertices} are the vertices of $X$ together with two new vertices $d_1,d_2$.
\item For every edge $(a,b\mid c)$ of $F$, the core vertex $c$ is joined to $a$ and to $b$; we call $\{a,b\}$ a \emph{partner pair}. There are no other edges between core vertices.
\item For every pair $\{u,w\}$ of non-adjacent core vertices that is not a partner pair, add two new \emph{private vertices} $p_{uw}$ and $q_{uw}$, each adjacent to $u$ and to $w$ and to no other core vertex.
\item Any two private vertices are adjacent.
\end{itemize}
The graph $G$ has $O(|X|^2)$ vertices and can be constructed in polynomial time. Since $F$ has girth at least four, two edges of $F$ share at most one vertex, so each pair of vertices lies in at most one edge of $F$ and each partner pair arises from exactly one edge.

\begin{claim}\label{cl:partner}
For every edge $(a,b\mid c)$ of $F$ we have $a\not\sim b$ and $C(a,b)=\{c\}$.
\end{claim}

\begin{proof}
Let $e=\{a,b,c\}$. Two core vertices are adjacent only if they lie in a common edge of $F$ and one of them is its centre. If $a\sim b$, then some edge $e'$ of $F$ contains $a$ and $b$, and $e'\ne e$ because neither $a$ nor $b$ is the centre of $e$; then $e$ and $e'$ share two vertices, a Berge cycle of length two. Hence $a\not\sim b$.

The vertex $c$ is a common neighbour of $a$ and $b$ by construction. Let $z\ne c$ be another common neighbour. It is a core vertex, since $\{a,b\}$ receives no private vertices. Core edges join vertices of a common edge of $F$, so $F$ has edges $E_1\supseteq\{a,z\}$ and $E_2\supseteq\{z,b\}$, both different from $e$ since $z\notin e$. If $E_1=E_2$, then $E_1$ and $e$ share $a$ and $b$ and form a Berge cycle of length two; otherwise $e,E_1,E_2$ form a Berge cycle of length three through $a$, $z$, $b$. Both contradict girth at least four, so $C(a,b)=\{c\}$.
\end{proof}

\begin{claim}\label{cl:diam}
Every core vertex has a private neighbour, and $G$ has diameter two.
\end{claim}

\begin{proof}
The vertices $d_1$ and $d_2$ have no core neighbours and lie in no partner pair. Hence, for every core vertex $x\ne d_1$, the pair $\{x,d_1\}$ receives private vertices, which are neighbours of both $x$ and $d_1$. In particular every core vertex has a private neighbour (for $d_1$, take $x=d_2$).

Now let $u,w$ be distinct non-adjacent vertices. If both are core vertices, they have a common neighbour: if $\{u,w\}$ is a partner pair, coming from an edge $(u,w\mid c)$ of $F$, then $c$ is adjacent to both by construction; otherwise $p_{uw}$ is adjacent to both. If $u$ is a core vertex and $w$ is a private vertex, then any private neighbour of $u$ is adjacent to $w$, because private vertices are pairwise adjacent (it differs from $w$ because $w\not\sim u$). Two private vertices are adjacent. So any two non-adjacent vertices have a common neighbour; since $G$ is not complete ($d_1\not\sim d_2$), it has diameter two.
\end{proof}

\begin{claim}\label{cl:equiv}
$F$ is $2$-colourable if and only if $\cmu(G)\le2$.
\end{claim}

\begin{proof}
By Lemma~\ref{lem:reform} and Claim~\ref{cl:diam}, it suffices to show that $F$ is $2$-colourable if and only if $\cH(G)$ is.

Suppose that $\cH(G)$ has a $2$-colouring. For every edge $(a,b\mid c)$ of $F$, Claim~\ref{cl:partner} shows that $\{a,b\}\cup C(a,b)=\{a,b,c\}$ is an edge of $\cH(G)$, so it is not monochromatic. Hence the restriction of the colouring to $X$ is a $2$-colouring of $F$.

Conversely, let $\psi$ be a $2$-colouring of $F$. Colour every $v\in X$ with $\psi(v)$, colour $d_1$ and $d_2$ arbitrarily, every vertex $p_{uw}$ red and every vertex $q_{uw}$ blue. We check that no edge $\{u,w\}\cup C(u,w)$ of $\cH(G)$ is monochromatic, according to the type of the non-adjacent pair $u,w$.
\begin{itemize}
\item If $\{u,w\}=\{a,b\}$ is a partner pair from the edge $(a,b\mid c)$, the edge of $\cH(G)$ is $\{a,b,c\}$ by Claim~\ref{cl:partner}, which is not monochromatic under $\psi$.
\item If $u,w$ are core vertices forming no partner pair, the edge contains $p_{uw}$, which is red, and $q_{uw}$, which is blue.
\item If $u$ is a core vertex and $w$ a private vertex, the edge contains every private neighbour of $u$ (see the proof of Claim~\ref{cl:diam}). By Claim~\ref{cl:diam}, $u$ has private neighbours, and they come in pairs $p_{uy},q_{uy}$, so the edge contains a red and a blue vertex.
\end{itemize}
Two private vertices are adjacent, so there are no other non-adjacent pairs. Hence $\cH(G)$ is $2$-colourable.
\end{proof}

\begin{proof}[Proof of Theorem~\ref{thm:main}]
The problem is in NP: a colouring with two colours can be checked in polynomial time by checking that any two non-adjacent vertices with the same colour have a common neighbour with the other colour. By Lemma~\ref{lem:girth}, $2$-colourability of $3$-uniform hypergraphs of girth at least four is NP-complete, and by Claims~\ref{cl:diam} and~\ref{cl:equiv}, $F\mapsto G(F)$ is a polynomial-time reduction from this problem to deciding $\cmu(G)\le2$ for graphs $G$ of diameter two.
\end{proof}

\section{Larger diameter}\label{sec:diamd}

Let $pq$ be an edge of a graph $H$. We say that a graph $H'$ is obtained from $H$ by \emph{attaching at the edge $pq$} if $H'$ is obtained from $H$ by adding new vertices and new edges, each new edge joining two new vertices or a new vertex to $p$ or $q$.

\begin{lemma}\label{lem:attach}
Let $H'$ be obtained from a connected graph $H$ by attaching at an edge $pq$. Then every shortest path in $H'$ between two vertices of $H$ lies in $H$; in particular, distances between vertices of $H$ are the same in $H$ and $H'$. Consequently, if $V(H')$ is partitioned into two MV sets of $H'$, then their intersections with $V(H)$ are MV sets of $H$.
\end{lemma}

\begin{proof}
A path in $H'$ between two vertices of $H$ that visits a new vertex must leave $H$ at one of $p,q$ and return at the other, since new vertices have no other neighbours in $H$. Replacing the part between $p$ and $q$, which has length at least two, by the edge $pq$ gives a shorter path. Hence every shortest path between vertices of $H$ lies in $H$, and distances between vertices of $H$ are the same in $H$ and $H'$.

For the consequence, let $u,w\in V(H)$ lie in the same set $S$ of the partition. Some shortest $u,w$-path in $H'$ has no internal vertex in $S$; it lies in $H$ and is a shortest $u,w$-path there, so $u$ and $w$ are $(S\cap V(H))$-visible in $H$.
\end{proof}

\medskip\noindent\textbf{The construction.} Let $d\ge3$ be the target diameter, and put $m=d-2\ge1$. Let $F$ be a $3$-uniform hypergraph of girth at least four; adding an isolated vertex if necessary, we may assume that $F$ has at least one vertex. Let $G=G(F)$ be the graph of Section~\ref{sec:reduction}. Let $p=p_{d_1d_2}$ and $q=q_{d_1d_2}$ be the private vertices of the pair $\{d_1,d_2\}$ (see the proof of Claim~\ref{cl:diam}). They are adjacent, and they are \emph{twins}: each is adjacent to $d_1$, to $d_2$ and to all other private vertices, and to nothing else.

\begin{figure}[ht]
\centering
\begin{tikzpicture}[scale=0.95,
  v/.style={circle,draw,inner sep=1.6pt,minimum size=5pt},
  wv/.style={v,fill=white},
  rv/.style={v,fill=red!65},
  bv/.style={v,fill=blue!55}]
  \draw[rounded corners=10pt,thick] (-3.3,1.15) rectangle (3.3,4.45);
  \node at (-2.75,4.05) {$G$};
  \draw[fill=gray!12] (0,3.75) ellipse (1.8 and 0.38);
  \node[font=\footnotesize] at (0,3.75) {$X$ (vertices of $F$)};
  \draw[fill=gray!12] (0,2.65) ellipse (1.8 and 0.34);
  \node[font=\footnotesize] at (0,2.65) {other private vertices};
  \draw[gray] (-0.75,2.93) -- (-0.95,3.42); \draw[gray] (0,2.97) -- (0,3.39); \draw[gray] (0.75,2.93) -- (0.95,3.42);
  \node[wv,label=left:{$d_1$}] (d1) at (-2.5,2.4) {};
  \node[wv,label=right:{$d_2$}] (d2) at (2.5,2.4) {};
  \node[rv,label=left:{$p$}] (p) at (-0.7,1.6) {};
  \node[bv,label=right:{$q$}] (q) at (0.7,1.6) {};
  \draw (p) -- (q);
  \foreach \t in {d1,d2} { \draw (p) -- (\t); \draw (q) -- (\t); }
  \foreach \x in {-0.55,0.55} { \draw[gray] (p) -- (\x,2.34); \draw[gray] (q) -- (\x,2.34); }
  \node[rv,label=left:{$r_1$}] (r1) at (-0.7,0.4) {};
  \node[bv,label=right:{$b_1$}] (b1) at (0.7,0.4) {};
  \node[rv,label=left:{$r_2$}] (r2) at (-0.7,-0.8) {};
  \node[bv,label=right:{$b_2$}] (b2) at (0.7,-0.8) {};
  \node (dots) at (0,-1.65) {$\vdots$};
  \node[rv,label=left:{$r_m$}] (rk) at (-0.7,-2.5) {};
  \node[bv,label=right:{$b_m$}] (bk) at (0.7,-2.5) {};
  \foreach \a/\b in {p/r1,p/b1,q/r1,q/b1,r1/r2,r1/b2,b1/r2,b1/b2} \draw (\a) -- (\b);
  \draw (r1) -- (b1); \draw (r2) -- (b2); \draw (rk) -- (bk);
  \draw (r2) -- (-0.7,-1.3); \draw (b2) -- (0.7,-1.3); \draw (r2) -- (-0.2,-1.3); \draw (b2) -- (0.2,-1.3);
  \draw (rk) -- (-0.7,-2.0); \draw (bk) -- (0.7,-2.0); \draw (rk) -- (-0.2,-2.0); \draw (bk) -- (0.2,-2.0);
  \node[font=\footnotesize] at (2.4,0.4) {$L_1$};
  \node[font=\footnotesize] at (2.4,-0.8) {$L_2$};
  \node[font=\footnotesize] at (2.4,-2.5) {$L_m$};
\end{tikzpicture}
\caption{The graph $G_d$, $m=d-2$: a ladder of levels $L_i=\{r_i,b_i\}$ attached to $G$ at the twin private vertices $p=p_{d_1d_2}$ and $q=q_{d_1d_2}$. Each level is an edge, consecutive levels are completely joined, and $L_1$ is completely joined to $\{p,q\}$. In $G$, the vertices $p$ and $q$ are adjacent to $d_1$, $d_2$ and all other private vertices, which form a clique; the vertices of $X$ are at distance two from $p$ and $q$. The colours are those used in Claim~\ref{cl:colour}.}
\label{fig:ladder}
\end{figure}
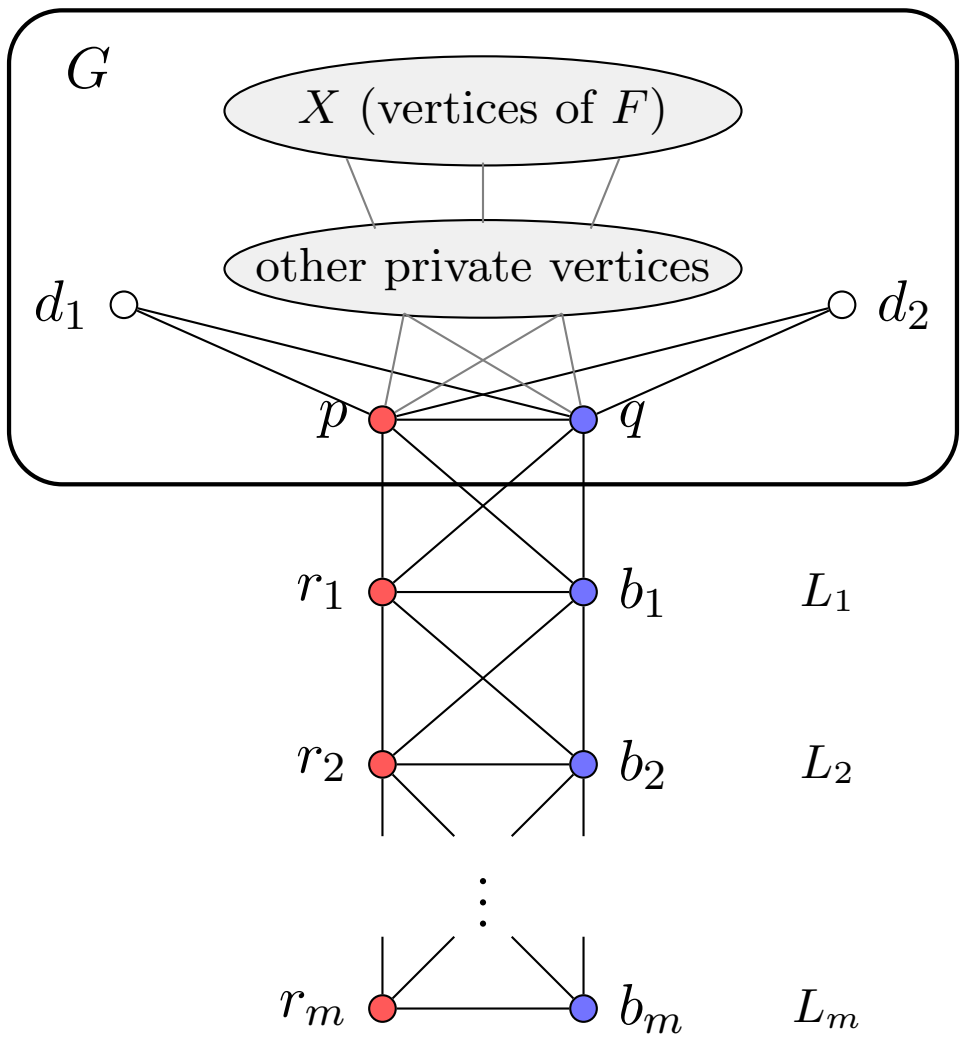

Let $G_d$ be obtained from $G$ by adding new vertices $r_i,b_i$ for $1\le i\le m$, with the edges $r_ib_i$ ($1\le i\le m$), all four edges between $L_i=\{r_i,b_i\}$ and $L_{i+1}$ ($1\le i<m$), and all four edges between $L_1$ and $\{p,q\}$; see Figure~\ref{fig:ladder}. We call $L_i$ the $i$-th \emph{level}. The construction is polynomial. Thus $G_d$ is obtained from $G$ by attaching at the edge $pq$, so Lemma~\ref{lem:attach} applies with $H=G$.

\begin{claim}\label{cl:dist}
$G_d$ has diameter $d$. Moreover, for $x\in L_i$ we have $d(x,p)=d(x,q)=i$ and $d(x,y)=i+d(q,y)$ for $y\in V(G)\setminus\{p,q\}$, where $d(q,y)=1$ if $y\in\{d_1,d_2\}$ or $y$ is private, and $d(q,y)=2$ if $y\in X$.
\end{claim}

\begin{proof}
By Lemma~\ref{lem:attach}, any two vertices of $G$ are at distance at most two in $G_d$. In $G$, the neighbours of $q$ are $d_1$, $d_2$ and the other private vertices, while each vertex of $X$ is non-adjacent to $q$ and hence at distance two from $q$, as $G$ has diameter two; the same holds for $p$, since $p$ and $q$ are twins. This gives the stated values of $d(q,y)$.

Regard $V(G)$ as level $0$. Every edge of $G_d$ joins vertices whose levels differ by at most one, so a path from $x\in L_i$ to a vertex of $G$ has at least $i-1$ edges before it reaches $L_1$, and then enters $G$ at $p$ or $q$. Hence $d(x,p)=d(x,q)=i$, and $d(x,y)=i+d(q,y)$ for $y\in V(G)\setminus\{p,q\}$; the path $x\,b_{i-1}\cdots b_1\,q$ followed by a shortest $q,y$-path attains these values. Similarly, two vertices in levels $i<j$ are at distance $j-i$.

So every distance in $G_d$ is at most $m+2$, with equality for $x\in L_m$ and $y\in X$ ($X$ is non-empty because $F$ has at least one vertex). Hence $G_d$ has diameter $m+2=d$.
\end{proof}

\begin{claim}\label{cl:back}
If $\cmu(G_d)\le2$, then $F$ is $2$-colourable.
\end{claim}

\begin{proof}
By Lemma~\ref{lem:attach}, a partition of $V(G_d)$ into at most two MV sets restricts to one of $V(G)$, so $\cmu(G)\le2$, and $F$ is $2$-colourable by Claim~\ref{cl:equiv}.
\end{proof}

\begin{claim}\label{cl:colour}
If $F$ is $2$-colourable, then $\cmu(G_d)\le2$.
\end{claim}

\begin{proof}
Colour $G$ as in the proof of Claim~\ref{cl:equiv} (so $p$ is red and $q$ is blue), and colour every $r_i$ red and every $b_i$ blue. Let $x,y$ be non-adjacent vertices of the same colour. The argument below uses only that each level, the pair $\{p,q\}$ and each pair $\{p_{yz},q_{yz}\}$ contain one red and one blue vertex; so by symmetry we may assume that $x$ and $y$ are red, and we find a shortest $x,y$-path whose internal vertices are all blue. If exactly one of $x,y$ lies in a level, let it be $x$; distances are given by Claim~\ref{cl:dist}.
The possibilities for $x$ and $y$ are the following.
\begin{enumerate}[label=(\arabic*)]
\item $x,y\in V(G)$: the colouring of $G$ is valid by Claim~\ref{cl:equiv}, and by Lemma~\ref{lem:attach} a shortest $x,y$-path in $G$ is also one in $G_d$.
\item $x=r_i$ and $y=r_j$ with $i<j$: the path $r_i\,b_{i+1}\cdots b_{j-1}\,r_j$ has length $j-i$.
\item $x=r_i$ and $y=p$ (the only red vertex of $\{p,q\}$; here $i\ge2$): the path $r_i\,b_{i-1}\cdots b_1\,p$ has length $i$.
\item $x=r_i$ and $y\in V(G)\setminus\{p,q\}$ with $y\sim q$: the path $r_i\,b_{i-1}\cdots b_1\,q\,y$ has length $i+1$.
\item $x=r_i$ and $y\in V(G)\setminus\{p,q\}$ with $y\not\sim q$: then $y\in X$. By Claim~\ref{cl:diam} it has a blue private neighbour $q_{yz}$, which is adjacent to $q$ (private vertices are pairwise adjacent, and $q_{yz}\ne q$ as $y\not\sim q$). The path $r_i\,b_{i-1}\cdots b_1\,q\,q_{yz}\,y$ has length $i+2$.
\end{enumerate}
In each case the length equals $d(x,y)$, so the path is a shortest path, and its internal vertices are blue. Hence both colour classes are MV sets of $G_d$.
\end{proof}

\begin{proof}[Proof of Theorem~\ref{thm:diamd}]
The case $d=2$ is Theorem~\ref{thm:main}. The problem is in NP for every $d$: given a colouring with two colour classes, two vertices $x,y$ of the same class $S$ are $S$-visible if and only if their distance in the subgraph induced by $(V\setminus S)\cup\{x,y\}$ equals their distance in the whole graph, which can be checked by breadth-first search. For $d\ge3$, Claims~\ref{cl:dist}, \ref{cl:back} and~\ref{cl:colour} show that $F\mapsto G_d$ is a polynomial-time reduction from the problem of Lemma~\ref{lem:girth} to deciding $\cmu(G)\le2$ for graphs $G$ of diameter $d$.
\end{proof}

\begin{remark}
Instead of $p_{d_1d_2},q_{d_1d_2}$, any pair $p_{uw},q_{uw}$ of private vertices can be used, that is, $\{u,w\}$ can be any non-adjacent pair of core vertices that is not a partner pair. The proofs are the same, with $X$ replaced by the set of core vertices other than $u$ and $w$.
\end{remark}

\section{More colours}\label{sec:kcol}

The arguments above extend from two colours to any fixed number of colours. A hypergraph is \emph{$k$-colourable} if its vertices can be coloured with $k$ colours so that no edge is monochromatic.

\begin{theorem}\label{thm:kcol}
For all fixed integers $k\ge2$ and $d\ge2$, the following problem is NP-complete: given a graph $G$ of diameter $d$, decide whether $\cmu(G)\le k$.
\end{theorem}

The case $k=2$ is Theorem~\ref{thm:diamd}. Membership in NP is shown as there. The proof for $k\ge3$ follows the same steps (the hypergraph reformulation, the hardness of colouring $3$-uniform hypergraphs of girth at least four, and the constructions $G(F)$ and $G_d$); we describe the changes.

\begin{lemma}\label{lem:reform-k}
Let $G$ be a graph of diameter two and $k\ge1$. Then $\cmu(G)\le k$ if and only if $\cH(G)$ is $k$-colourable.
\end{lemma}

\begin{proof}
As in the proof of Lemma~\ref{lem:reform}, a partition of $V(G)$ into at most $k$ MV sets is a colouring with $k$ colours whose colour classes are MV sets, and a colour class is an MV set if and only if any two non-adjacent vertices $u,w$ in it have a common neighbour of another colour, that is, if and only if no set $\{u,w\}\cup C(u,w)$ with $u\not\sim w$ is monochromatic in that colour.
\end{proof}

\begin{lemma}\label{lem:3unif-k}
For every fixed $k\ge3$, it is NP-complete to decide whether a $3$-uniform hypergraph is $k$-colourable.
\end{lemma}

\begin{proof}
Membership in NP is clear. We reduce from graph $k$-colourability, which is NP-complete~\cite[GT4]{GJ1979}. Given a graph $G$, let $W=\{w_1,\dots,w_{2k}\}$ be a set of $2k$ new vertices, and let $H$ be the $3$-uniform hypergraph with vertex set $V(G)\cup W$ whose edges are all $3$-element subsets of $W$, together with the sets $\{u,v,w\}$ for $uv\in E(G)$ and $w\in W$. It can be constructed in polynomial time.

In a $k$-colouring of $H$, no three vertices of $W$ have the same colour, since they form an edge. As $|W|=2k$, every one of the $k$ colours therefore occurs on exactly two vertices of $W$.

Let $c$ be a proper $k$-colouring of $G$. Colour $V(G)$ by $c$, and give $w_{2j-1}$ and $w_{2j}$ colour $j$ for $1\le j\le k$. Every $3$-element subset of $W$ contains two colours, and every edge $\{u,v,w\}$ with $uv\in E(G)$ contains the two colours $c(u)\ne c(v)$. So $H$ is $k$-colourable.

Conversely, take a $k$-colouring of $H$. If $uv\in E(G)$ and $u,v$ have the same colour $\gamma$, choose $w\in W$ of colour $\gamma$, which exists by the previous paragraph; then $\{u,v,w\}$ is monochromatic, a contradiction. So the colouring restricted to $V(G)$ is a proper $k$-colouring of $G$.
\end{proof}

\begin{lemma}\label{lem:girth-k}
For every fixed $k\ge2$, it is NP-complete to decide whether a $3$-uniform hypergraph of girth at least four is $k$-colourable.
\end{lemma}

\begin{proof}
For $k=2$ this is Lemma~\ref{lem:girth}; let $k\ge3$. We repeat the proof of Lemma~\ref{lem:girth} with the following changes. Applying the theorem of Erd\H{o}s and Hajnal~\cite[Corollary~13.4]{EH1966} with $r=k+1$ instead of $r=3$ (and with edge size $3$ and $s=3$ as before) gives a $3$-uniform hypergraph that is not $k$-colourable and in which any $t\le3$ edges cover at least $2t+1$ vertices; it has girth at least four, as before. Let $H_0$ be an edge-minimal non-$k$-colourable sub-hypergraph of it, fix an edge $e_0=\{x,y,z\}$ of $H_0$, and let $D=H_0-e_0$. Then (a) $D$ is $k$-colourable; (b) in every $k$-colouring of $D$ the vertices $x,y,z$ have the same colour, since otherwise the colouring would be a $k$-colouring of $H_0$, and permuting the colours, $D$ has $k$-colourings in which $x,y,z$ all have any prescribed colour; and (c) no edge of $D$ contains two of $x,y,z$. We reduce from the problem of Lemma~\ref{lem:3unif-k}. The construction of $F'$, the proof that $F$ is $k$-colourable if and only if $F'$ is, and the proof that $F'$ has girth at least four are then word for word as before, with $2$-colourings replaced by $k$-colourings.
\end{proof}

\begin{proof}[Proof of Theorem~\ref{thm:kcol} for $k\ge3$]
Let $F$ be a $3$-uniform hypergraph of girth at least four, with at least one vertex as in Section~\ref{sec:diamd}, and let $G(F)$ and, for $d\ge3$, $G_d$ be the graphs of Sections~\ref{sec:reduction} and~\ref{sec:diamd}. By Claims~\ref{cl:diam} and~\ref{cl:dist}, which do not involve colours, they have diameter two and $d$, respectively. By Lemma~\ref{lem:girth-k}, it suffices to show that $F$ is $k$-colourable if and only if $\cmu(G(F))\le k$, and if and only if $\cmu(G_d)\le k$.

\emph{The graph $G=G(F)$.} By Lemma~\ref{lem:reform-k} it suffices to show that $F$ is $k$-colourable if and only if $\cH(G)$ is. If $\cH(G)$ is $k$-colourable, then by Claim~\ref{cl:partner} every edge of $F$ is an edge of $\cH(G)$, so the restriction of the colouring to $X$ is a $k$-colouring of $F$. Conversely, given a $k$-colouring $\psi$ of $F$, colour every $v\in X$ with $\psi(v)$, colour $d_1$ and $d_2$ arbitrarily, every vertex $p_{uw}$ with colour $1$ and every vertex $q_{uw}$ with colour $2$. The proof of Claim~\ref{cl:equiv} applies verbatim, with red and blue read as colours $1$ and $2$: every edge of $\cH(G)$ is either an edge of $F$ or contains a vertex of colour $1$ and a vertex of colour $2$.

\emph{The graph $G_d$, $d\ge3$.} If $V(G_d)$ is partitioned into at most $k$ MV sets of $G_d$, then their intersections with $V(G)$ are MV sets of $G$; the proof of Lemma~\ref{lem:attach} applies verbatim to partitions into any number of MV sets. So $\cmu(G)\le k$, and $F$ is $k$-colourable by the first part. Conversely, given a $k$-colouring of $F$, colour $G$ as in the first part (so $p$ has colour $1$ and $q$ has colour $2$), and give every $r_i$ colour $1$ and every $b_i$ colour $2$. Let $x,y$ be non-adjacent vertices of the same colour $\gamma$. If $\gamma\in\{1,2\}$, the proof of Claim~\ref{cl:colour} applies, with red and blue read as colours $1$ and $2$ and with Claim~\ref{cl:equiv} replaced by the first part of this proof. If $\gamma\notin\{1,2\}$, then $x,y\in V(G)$, as the ladder uses only colours $1$ and $2$, and case~(1) of that proof applies, with the same replacement. Hence all colour classes are MV sets of $G_d$, and $\cmu(G_d)\le k$.
\end{proof}

\section*{Use of generative AI}
During this work the author used Claude Opus 5.5 (Anthropic), which proposed the reduction and the proofs in this paper and helped prepare the manuscript. The author directed the investigation through questions that shaped its course, verified every step of each proof by hand, and takes full responsibility for the content of the publication.

\end{document}